\documentclass[aps,prb,showpacs,twocolumn,groupedaddress]{revtex4-1}

\usepackage{graphicx}
\usepackage{epstopdf}
\usepackage{hyperref}%

\begin{document}


\title{Scaling behavior of the critical current in clean epitaxial Ba(Fe$_{1-x}$Co$_x$)$_2$As$_2$ thin films}


\author{K.\,Iida}
\email[Electronical address:\,]{k.iida@ifw-dresden.de}
\author{J.\,H\"{a}nisch}
\author{T.\,Thersleff}
\author{F.\,Kurth}
\author{M.\,Kidszun}
\author{S.\,Haindl}
\author{R.\,H\"{u}hne}
\author{L.\,Schultz}
\author{B.\,Holzapfel}
\affiliation{IFW Dresden, P.\,O.\,Box 270116, D--01171 Dresden, Germany}


\date{\today}

\begin{abstract}
The angular-dependent critical current density, $J_{\rm c}(\Theta)$, and the upper critical field, $H_{\rm c2}(\Theta)$, of epitaxial Ba(Fe$_{1-x}$Co$_x$)$_2$As$_2$ thin films have been investigated. No $J_{\rm c}(\Theta)$ peaks for $H\parallel c$ were observed regardless of temperatures and magnetic fields. In contrast, $J_{\rm c}(\Theta)$ showed a broad maximum at $\Theta=90^{\circ}$, which arises from intrinsic pinning. All data except at $\Theta=90^{\circ}$ can be scaled by the Blatter plot. $H_{\rm c2}(\Theta)$ near $T_{\rm c}$ follows the anisotropic Ginzburg-Landau expression. The mass anisotropy increased from 1.5 to 2 with increasing temperature, which is an evidence for multi-band superconductivity.
\end{abstract}


\maketitle

Tremendous efforts on the fabrication of superconducting iron pnictide thin films involving fluorine-doped (Rare-earth elements)FeAsO,\cite{01,02,03,04} Co-containing (alkaline elements)Fe$_{2}$As$_{2}$,\cite{05,06,07,08,09,10,11} FeSe\cite{12} and Te-containing FeSe\cite{13} have been made for both investigating their intrinsic properties and exploring possible device applications. For such investigations, Co-containing BaFe$_{2}$As$_{2}$ (hereafter, Ba-122) films are, perhaps, the most feasible with regard to their easy growth by pulsed laser deposition (PLD) and, most importantly, their chemical stability against ambient atmosphere.\cite{08}

Recent investigations on this material revealed numerous similarities to the so-called 123-cuprates in terms of physical parameters albeit a large difference in the superconducting transition temperature, $T_{\rm c}$, of $\approx$70\,K. YBa$_2$Cu$_3$O$_7$ (Y-123) and Co-containing Ba-122, for example, have a similar in-plane coherence length, $\xi_{ab}$, of around 2\,nm at low temperatures, which originates from both their small Fermi velocity and low carrier concentration.\cite{14} Therefore, grain boundaries (GBs), even with small misorientation angles of 3-5$^{\circ}$, are predicted to become a critical issue for multiband Co-containing Ba-122 system, and indeed the current-limiting effect of GBs has been reported.\cite{09}

In this Letter, we report on the transport properties of Ba(Fe$_{1-x}$Co$_x$)$_2$As$_2$ thin films on (La,Sr)(Ta,Al)O$_3$ (LSAT) substrates and evaluate their pinning performance. In a previous publication,\cite{11} the films on SrTiO$_3$ (STO) had a better epitaxial quality with a higher $T_{\rm c}$ compared to those obtained on LSAT. However, the STO substrate becomes electrically conducting under UHV conditions,\cite{15} presenting an obstacle to transport measurements particularly near the normal-superconducting transition. Therefore, we have selected the LSAT substrate in the present investigation.

High-quality epitaxial Co-containing Ba-122 thin films of 30\,nm thickness have been deposited on (100)-oriented cubic LSAT substrates by PLD, where the BaFe$_{1.8}$Co$_{0.2}$As$_2$ sintered target was ablated by 248\,nm KrF radiation at a frequency of 5\,Hz, an energy density of $3-5\,\rm Jcm^{-2}$, and a base pressure of 10$^{-9}$\,mbar. Co-containing Ba-122 films seem to be weakly volatile against ambient atmosphere albeit they have been reported to be stable.\cite{08} Therefore,  after the deposition, a Cr cap layer of 5\,nm was deposited on the resulting films at room temperature in order to prevent any degradation of superconducting properties during measurements. A detailed description of the film preparation can be found in ref.\cite{11}.

\begin{figure}[t]
	\centering
		\includegraphics[height=8cm,width=7.5cm]{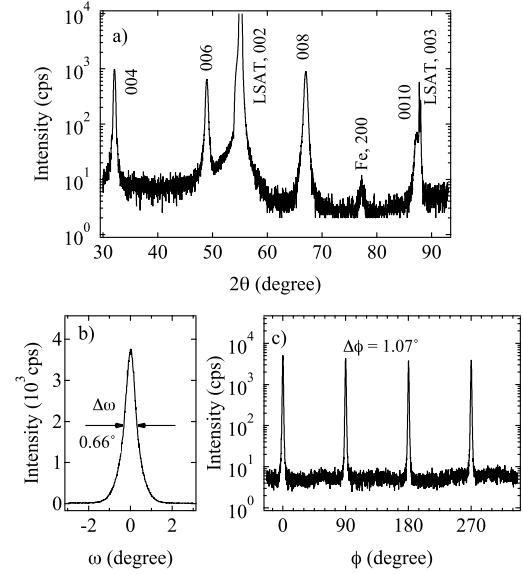}
		\caption{(a) The film has been grown with $c$-axis texture. The 200 reflection of Fe was observed at $2\rm\theta\approx77^{\circ}$. (b) The FWHM value, $\Delta\omega$, was 0.66$^{\circ}$ from the 004 rocking curve.(c) Only a sharp peak at every 90$^{\circ}$ was observed in the 103 reflection of $\phi$-scans. The average FWHM value, $\rm \Delta\phi$, of the peaks was 1.07$^{\circ}$. These results prove that the resulting film was of good crystalline quality.} 
\label{fig:figure1}
\end{figure}

\begin{figure}[t]
	\centering
			\includegraphics[height=12cm,width=7.8cm]{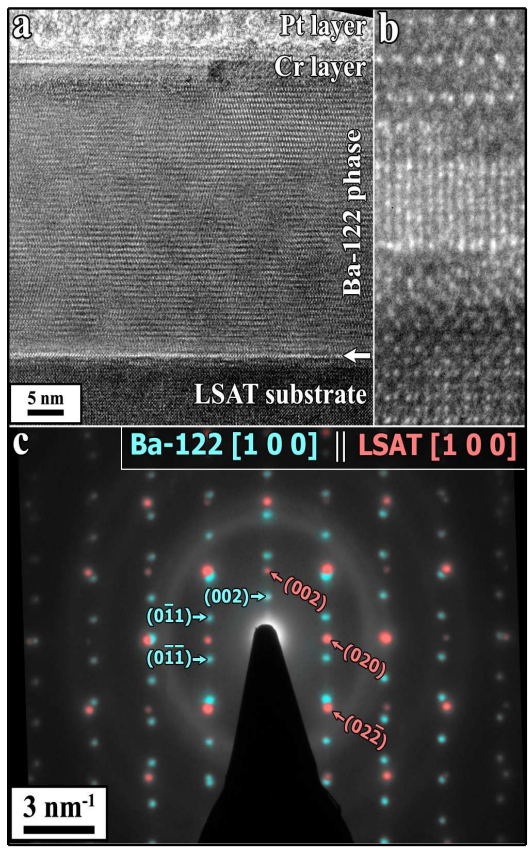}
		\caption{(Color online) (a) Cross-sectional TEM image of the Co-containing Ba-122 film. No misoriented grains were observed. The image was acquired with the beam parallel to LSAT [100]. An arrow points to the interfacial region. (b) HRTEM image of the interfacial region acquired under the same conditions as (a). A 2\,nm thick layer of Fe in the [110] zone axis can be seen. This image is exactly 3\,nm wide. (c) SAED pattern of the region in (a) revealing the epitaxial relationship between the LSAT substrate and the Co-containing Ba-122 layer. The pattern has been artificially colored to clarify the substrate and film reflections. The lowest order reflections have been indexed. The observed rings correspond to amorphous Pt in the protective surface layer.} 
\label{fig:figure2}
\end{figure}

For transport measurements, the films were cut into slabs measuring 1-2\,mm in width and 5-8\,mm in length with a razor blade. Silver paint was employed for electrical contacts. Superconducting properties were measured in a commercial Physical Property Measurement System (PPMS, Quantum Design) by a standard four-probe method with a criterion of 1\,$\rm\mu Vcm^{-1}$ for evaluating $J_{\rm c}$. In the angular-dependent $J_{\rm c}$ measurements, the magnetic field, $H$, was applied in the maximum Lorentz force configuration ($H$ perpendicular to $J$) at an angle of $\Theta$ measured from the $c$-axis. The same configuration was employed for the angular-and field-dependent resistance measurements near $T_{\rm c}$ to evaluate the upper critical field, $H_{\rm c2}$.

The crystalline quality and phase purity of the thin film was examined prior to detailed transport measurements by $\theta\rm/2\theta$\,-\,scans, 103 $\phi$\,-\,scans, and $\omega$\,-\,scans of the 004 diffraction peak. It is apparent from fig.\,\ref{fig:figure1}\,(a) that only 00$l$ peaks of the Co-containing Ba-122 thin film together with the 200 reflection of Fe were observed. The full width at half maximum (FWHM), $\Delta\omega$, of the 004 rocking curve was 0.66$^{\circ}$, as exhibited in fig.\,\ref{fig:figure1}\,(b). Furthermore, no additional peaks in the $\phi$-scans of the 103 reflection were observed outside of a sharp, strong peak at every 90$^{\circ}$, as shown in fig.\,\ref{fig:figure1}\,(c). The average FWHM, $\Delta\phi$, of the peaks was 1.07$^{\circ}$. A further investigation on the crystalline quality of the film has been performed by transimission electron microscopy, TEM. Shown in figs.\,\ref{fig:figure2}\,(a) and (b) are the cross-sectional TEM images together with its selected area of the electron diffraction (SAED) pattern (c), taken in a ${\rm C_s}$-corrected 300\,kV, FEI Titan. Fig.\,\ref{fig:figure2}\,(a), an overview micrograph representative of the entire 10\,$\rm \mu m$ long TEM lamella surveyed, reveals neither appreciable defects nor any grain boudaries. It is further apparent from figs.\,\ref{fig:figure2}\,(c) and \,\ref{fig:figure1}\,(c) that the film grows highly epitaxially with the relation (001)[100]Ba-122$\|$(001)[100]LSAT without any misoriented grains, which was also confirmed from the overall TEM image and pole figure measurements. From these results, we conclude that the Co-containing Ba-122 thin film is clean and of excellent crystalline quality without grain boundaries.

The high-resolution TEM image at the interface between the film and substrate reveals a thin epitaxial Fe layer of $\approx2$\,nm (fig.\,\ref{fig:figure2}\,(b)). While Fe was not detected in the PLD target material by powder X-ray diffraction, the 200 reflection of Fe is commonly observed in thin films rather than high-intensity peaks such as 110 or 210, as shown in fig.\,\ref{fig:figure1}\,(a).\cite{05,08,10} These results suggest that Fe is textured and a by-product formed during the deposition. From the above results, the epitaxial relation of the resulting film is (001)[100]Ba-122$\|$(001)[110]Fe$\|$(001)[100]LSAT, which was also confirmed by $\phi$-scans. Here, the lattice constant $a$ of Fe (0.286\,nm) multiplied by $\sqrt2$ is close to those of LSAT (0.387\,nm) and BaFe$_{1.8}$Co$_{0.2}$As$_{2}$ (0.396\,nm).

\begin{figure}[b]
	\centering
		\includegraphics[height=11cm,width=7cm]{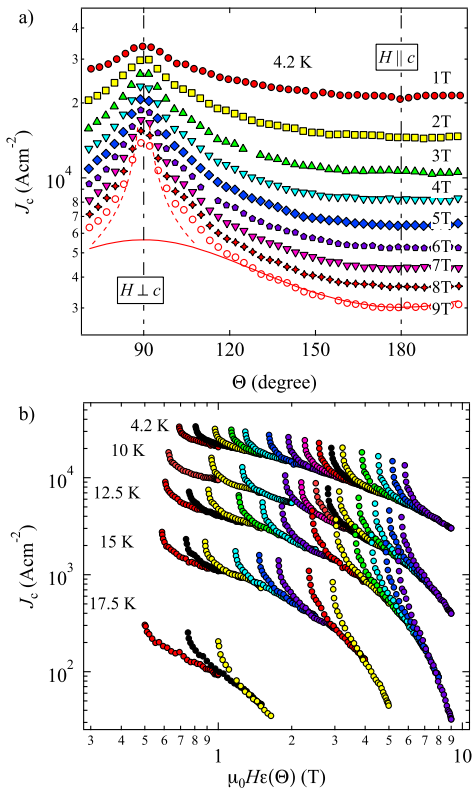}
		\caption{(Color online) (a) Angular dependence of $J_{\rm c}$ for the Ba(Fe$_1$$_-$$_x$Co$_x$)$_2$As$_2$ film at 4.2\,K under several magnetic fields. $J_{\rm c}$ exhibits a broad maximum centered at $H\perp c$. No additional peaks except at $H\perp c$ were observed. The solid and broken red lines represent the random defect and intrinsic contribution at 9\,T, respectively. (b) All $J_{\rm c}$ data can be scaled with an anisotropy parameter of $\gamma$ ranging from 1.5$\sim$2 at given temperatures except for the angular range close to $H\perp c$.} 
\label{fig:figure3}
\end{figure}

Transport $J_{\rm c}$ of the resulting thin film, which shows a $T_{\rm c,90}$ of 22.3\,K with a transition width, $\Delta$$T_{\rm c}$, of 1.8\,K ($\Delta$$T_{\rm c}$=$T_{\rm c,90}$-$T_{\rm c,10}$, where $T_{\rm c,90}$ and $T_{\rm c,10}$ are defined as  $90\%$ and $10\%$ of the normal resistivity at 25\,K), as a function of magnetic field at different temperatures has been evaluated. The magnetic field was applied parallel and perpendicular to the film $c$-axis. $J_{\rm c}$ values for $H$$\parallel$$c$ are always lower than for $H$$\perp$${c}$, indicating that the flux pinning is anisotropic. Further investigation of the anisotropy of flux pinning has been done by measuring $J_{\rm c}$ as a function of angle, $\Theta$, under different magnetic fields and temperatures. Shown in fig.\,\ref{fig:figure3}\,(a) is $J_{\rm c}(\Theta)$ under different magnetic fields at 4.2 K. $J_{\rm c}(\Theta)$ always has a broad maximum positioned at $\Theta$=90$^{\circ}$ ($H$\,$\perp$\,${c}$), and this peak becomes more prominent with increasing magnetic field. These features are typically observed in layered superconductors like Y-123 due to intrinsic pinning arising from correlated $ab$-planes and a modulation of the superconducting order parameter.\cite{16} Here we emphasize that no $J_{\rm c}$ peaks at $\Theta$=180$^{\circ}$ ($H\parallel c$) were observed in the whole range of temperatures as well as magnetic fields, which is consistent with the TEM results (i.e. no $c$-axis correlated defects were observed, as shown in fig\,\ref{fig:figure2}\,(a).). However, S.\,Lee \emph{et al.}\cite{10} have observed $c$-axis correlated defects on Ba-122 films, which led to enormous $J_{\rm c}$ peaks for $H\parallel c$. This discrepancy may originate from the different deposition condition such as temperature and PLD target composition. According to ref. \cite{17}, the Co content in Ba-122 significantly influences the superconducting properties. For instance, under-doped Ba(Fe$_1$$_-$$_x$Co$_x$)$_2$As$_2$ ($x$=0.054) single crystals show a large $J_{c}$ of over 10$^5$Acm$^{-2}$ at $T=0.7\,T_{\rm c}$ and $H_{\rm m}$ ($H_{\rm m}$ is the field of fishtail magnetization maximum in the hysteresis loop) due to structural domains, whereas the corresponding value of the over-doped crystals ($x$=0.086) exhibited one order magnitude lower values albeit the former crystal showed a lower $T_{\rm c}$ of 19.2\,K than the latter one ($T_{\rm c}$=20.5\,K). The film composition in this study may be located further in the over-doped region compared with the PLD target, since Fe has been consumed during film deposition.

In the absence of correlated defects, (i.e. if only the mass anisotropy is responsible for the $J_{\rm c}$ anisotropy) $J_{\rm c}$ can be scaled with $H_{\rm eff}$, $(H_{\rm eff}=H\epsilon(\Theta)$, $\epsilon(\Theta)=\sqrt{\cos ^2(\Theta)+\gamma^{-2}\sin ^2(\Theta)})$, where $\gamma$ is the mass anisotropy ratio.\cite{18} This was, for instance, applied to one-band Y-123 by L. Civale $et$ $al$\cite{19} and two-band MgB$_2$ superconducting films in the dirty limit by S. Sen $et$ $al$,\cite{21} however, $J_{\rm c}(\Theta)$ for the latter films can not be scaled below 33\,K. The scaling behavior of $J_{\rm c}(\Theta)$ as a function of $H_{\rm eff}$ at various temperatures is exhibited in fig.\,\ref{fig:figure3}\,(b). All data except in the vicinity of $H$$\perp$$c$ collapse onto a single curve with relatively small anisotropies, $\gamma$, of 1.5-2. This means that random defects are dominant in those regions, whereas a failure to scale the data close to $H$$\perp$$c$ is owing to the intrinsic pinning, which originates from the modulation of the superconducting order parameter along the $c$-axis. Here, the value of the out-of-plane coherence length $\xi_c$ for the Co-containing Ba-122 is almost equivalent to its lattice constant $c$. Once the $J_{\rm c}$($H_{\rm eff}$) has been obtained (i.e. collapsed data in fig.\,\ref{fig:figure3}\,(b)), $J_{\rm c}$ owing to random defects can be re-plotted in the angular dependent $J_{\rm c}$. The solid line in fig.\,\ref{fig:figure3}\,(a) is an example at 4.2\,K and $\mu_0H$=9\,T of $J_{\rm c}$ due to random defects.

On the other hand, $J_{\rm c}(\Theta)$ owing to the intrinsic pinning described above may be expressed by the Tachiki-Takahashi model (i.e. $J_{\rm c}(\Theta)$=$J_{\rm c}$(180$^{\circ}$)$\vert \rm cos(\Theta)\vert^{-0.5}$).\cite{20} It is clear that the measured $J_{\rm c}(\Theta)$ in the angular region close to 90$^{\circ}$ can be fitted by this model as shown in the fig.\,\ref{fig:figure3}\,(a) (see red broken lines).

However, the experimental data deviate from the calculation in the range of 100$^{\circ}$ to 110$^{\circ}$ (70$^{\circ}$ to 80$^{\circ}$) and this deviation becomes larger with increasing temperature or decreasing magnetic field. Plausible reasons for this deviation may originate from either a gradual breakdown of the Tachiki-Takahashi model due to the small anisotropy or the Pearl vortices due to a thinner film thickness than the penetration depth.\cite{24} 

\begin{figure}[t]
	\centering
		\includegraphics[height=11cm,width=7cm]{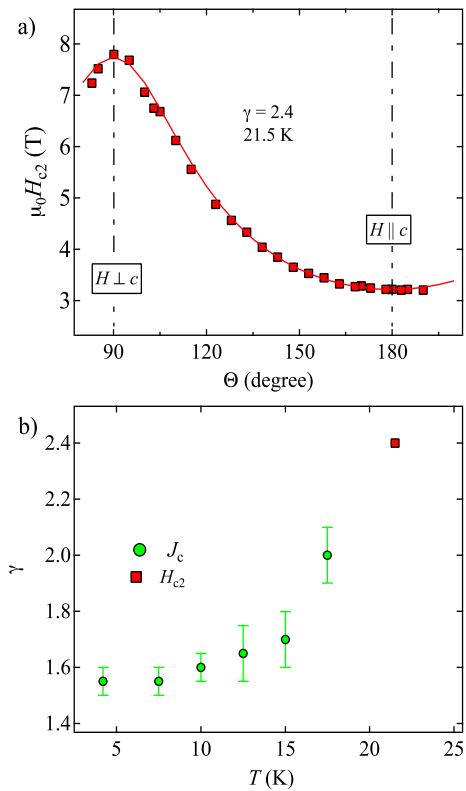}
		\caption{(Color online) (a) Angular dependence of $H_{\rm c2}$ at 21.5\,K for the Ba(Fe$_{1-x}$Co$_x$)$_2$As$_2$ thin film. The solid line represents the anisotropic GL equation with a $\gamma$ value of 2.4. (b) The $\gamma$ values obtained by two different methods described in the text were observed to increase with increasing temperatures.} 
\label{fig:figure4}
\end{figure}

The angular-dependent upper critical field, $H_{\rm c2}(\Theta)$, at 21.5\,K, shown in fig.\,\ref{fig:figure4}\,(a), was determined from the onset of superconductivity defined as a $90\%$ of the normal resistance at 25\,K. The resulting $H_{\rm c2}(\Theta)$ exhibited a maximum at $\Theta$=90$^{\circ}$$(H^{ab}_{\text{c2}})$ and a minimum at $\Theta$=180$^{\circ}$$(H^{c}_{\text{c2}})$. For comparison, we show a calculation based on the anisotropic Ginzburg-Landau (GL) theory, $H_{\rm c2}(\Theta)$=$H^{c}_{\text{c2}}$$(\rm cos^2(\Theta)+sin^2(\Theta)/\gamma^2)^{-0.5}$,with $\gamma$=$H^{ab}_{\text{c2}}/H^{c}_{\text{c2}}$. Clearly, the GL expression fits very well the measured $H_{\rm c2}(\Theta)$ similar to ref.\,\cite{23}, which is in contrast to the results of two-band MgB$_2$ superconductors in the dirty limit.\cite{22} Shown in fig.\,\ref{fig:figure4}\,(b) is the $\gamma$ values evaluated by two different methods, in which the Blatter scaling approach for the $J_{\rm c}(\Theta$,\,$H$) at low $T$, where the finite $J_{\rm c}$ can be measured and for the $H_{\rm c2}(\Theta)$ at high $T$, where $\mu_{0}H_{\rm c2}<9\,\rm T$, as a function of temperature. It is clear from fig.\,\ref{fig:figure4}\,(b) that $\gamma$ is increasing with increasing temperature, which is consistent with dominant inter-band scattering.

In summary, we have investigated transport properties of Co-containing Ba-122 thin films deposited on insulating LSAT substrates. TEM microstructural observation revealed that a thin epitaxial Fe layer was present at the film/substrate interface. Despite of the Fe layer between film and substrate, epitaxial growth with the relation (001)[100]Ba-122$\|$(001)[110]Fe$\|$(001)[100]LSAT was achieved. No peaks representing $c$-axis correlated pinning were recorded in the angular dependence of $J_{\rm c}$ in the whole range of temperatures and magnetic fields. On the other hand, $J_{\rm c}$ exhibits a broad maximum centered at $H\perp c$. All data except for those close to $H\perp c$ can be scaled with the mass anisotropy function $\epsilon(\Theta)$, as described by G. Blatter $et$ $al$. and L. Civale $et$ $al$. indicating a pinning contribution arising from uncorrelated defects. The extracted temperature dependence of the $\gamma$ value increases with increasing temperature, which is a strong evidence for multi-band superconductivity. Near $T_{\rm c}$, the angular dependence of the upper critical field $H_{\rm c2}(\Theta)$ is well fitted by the anisotropic Ginzburg-Landau theory with a mass anisotropy of 2.4, which follows well the temperature dependence $\gamma$ value by $J_{\rm c}(\Theta$) scaling. 

\begin{acknowledgments}
The authors would like to thank Drs.\,G.\,Fuchs and S.-L.\,Drechsler for fruitful discussions, J.\,Werner, M.\,Deutschmann, C.\,Nacke, K.\,Tscharntke, M.\,K\"{u}hnel and U.\,Besold for their technical support as well as O.\,Wilhelmi and FEI company for their assistance with the TEM lamella preparation.
\end{acknowledgments}

\bibliography{PRB-LM12513BR.bib}

\end{document}